\documentstyle[prl,aps,epsfig]{revtex}
\twocolumn
\begin{document}
\draft
\twocolumn[\hsize\textwidth\columnwidth\hsize\csname@twocolumnfalse\endcsname
\title{Phase Separation and Coarsening in
Electrostatically Driven Granular Media}
\author{I.S. Aranson$^{1}$, B. Meerson$^{1,2}$, P.V. Sasorov$^{3}$
and V.M. Vinokur$^{1}$}
\address{$^{1}$Argonne National Laboratory, 9700 S. Cass Avenue, Argonne, IL 60439}
\address{$^{2}$Racah Institute of Physics, Hebrew University of  Jerusalem, Jerusalem 91904, Israel}
\address{$^{3}$Institute of Theoretical and Experimental Physics, Moscow 117259, Russia}
\maketitle
\begin{abstract}
A continuum model for the phase separation and coarsening, observed in
electrostatically driven granular media, is formulated
in terms of a Ginzburg-Landau equation subject to
conservation of the total number of grains.
In the regime of well-developed clusters, the
continuum model is used to derive ``sharp-interface" equations that govern the dynamics of the interphase
boundary. The model captures the essential physics of this system.
\end{abstract}
\vskip1pc]
\narrowtext

Despite extensive work in the last two decades, the physics of granular flow is still poorly
understood, especially in the limit of strongly inelastic granular collisions when no
first-principles hydrodynamic description is available \cite{Jaeger,Kadanoff}. Additional
complications include contact interactions which become dominant when the grain size goes below
$0.1$ mm. As particles may acquire an electric charge, a new type of dynamics appears which is
governed by the interplay between long-range electromagnetic and short-range contact forces. In
this regime,
electrostatic excitation of granular media becomes possible.
It offers an
opportunity for an investigation of granular systems consisting of small particles.

Recently,
off-equilibrium phase
separation and coarsening in electrostatically driven granular submonolayers
was reported
  \cite{Aranson,Howell}. The inset of Fig. 1
shows a schematic of the setting. Conducting particles are placed between the plates of a plane
capacitor which is energized by a constant (DC) or alternating (AC) electric field $E$. When
particles are in contact with the capacitor plate, they acquire an electric charge. If the electric
field exceeds a critical value, the resulting (upward) electric force overcomes the gravity force
$mg$ ($g$ is gravity acceleration and $m$ is mass of particle) and pushes the particles upward.
When a particle hits the upper plate, it gets the opposite charge and falls back. Changing the
frequency $\omega$ of the  AC    field $E(t)=E_0 \cos(\omega t)$, one can control the particle
elevation and avoid the collisions with the upper plate.

It was found in Ref. \cite{Aranson} that the particles remain immobile on the bottom plate at
$E<E_1$ (the precipitate state). If the electric field is larger than a second threshold value,
$E_2>E_1$, the system is in a gas-like state. 
This second field $E_2$ is 50\%-70\% larger than $E_1$.
It was found that, upon decreasing the field below $E_2$ (in the interval
$E_1<E<E_2$) there is nucleation of precipitate, and small densely packed clusters form and start
to grow. The clusters then exhibit Ostwald ripening \cite{Ostwald}: smaller clusters shrink, while
larger clusters grow at their expense.
Molecular dynamics simulations showed qualitative agreement
with experimental results on phase separation and coarsening \cite{Aranson}.

In this Letter  we develop a continuum description of these phenomena. The  model captures the
essential phenomenology, correctly reproducing different possible morphologies of coarsening and
providing a correct quantitative description of its dynamic scaling behavior.

Let the granulate consist of $N \gg 1$ identical spherical particles with mass $m$ and radius
$\sigma$ (which is small compared with the plate spacing $h$). The continuum version of the model
employs two fields: the densities of the precipitate, $n({\mathbf r},t)$ and gas phase
$n_g({\mathbf r},t)$, where ${\mathbf r}=(x,y)$ and $x$ and $y$ are the horizontal coordinates. For
concreteness, we will consider a low-frequency electric field (so that the gas density is almost
independent of the vertical coordinate) and measure $n_g$ in cm$^{-2}$.

The phase separation is caused by electrostatic
screening \cite{Aranson}: a decrease in the vertical electric force $F$, exerted on a grain in contact with the
bottom plate, caused by the presence of other grains. Thus,
the $F(n)$ dependence is an important element of our model.
This dependence can be easily found in the dilute limit $n\sigma^2 \ll
1$. Consider two grains lying on the bottom plate the distance $L \gg \sigma$ apart. The second
grain and its mirror image in the plane form a dipole with the dipole moment of order of $\sigma^3
E$. This dipole produces an electric field at the location of the first grain: $\delta E \sim -
(\sigma/L)^3 E$. Therefore, the net force acting on the first grain is reduced and becomes $F_0
\left[1-\kappa (\sigma/L)^3)\right]$. If there are many grains lying far apart, we can simply sum
up these vertical forces and obtain the net force
\begin{equation}\label{homogen}
  F(n)=F_0 (1-k_1\, \sigma^3 n^{3/2})\,.
\end{equation}
Here $F_0=1.36\dots  \sigma^2 E^2$ is the vertical force exerted on a \textit{single} grain in
contact with the bottom plate \cite{Aranson}, and $\kappa, \, k_1 = {\cal O}(1)$ are numerical
factors.

Although no analytic expression for $F(n)$ is presently available for intermediate and large $n$,
it is clear that $F(n)$ should decrease with $n$ \cite{Aranson}. A decreasing $F(n)$ dependence
leads, at intermediate values of $E$, to a \textit{segregation instability} and
\textit{bistability}. Indeed, the density value $n=n_*$ such that $F(n_*)=mg$ is in an unstable
equilibrium. For $n<n_*$, $F(n)$ exceeds $mg$ so the particles ``evaporate" until the ``empty
state" $n=0$ is reached. If $n>n_*$, $F(n)<mg$ and the particles remain immobile. However, as gas
particles hit these regions and (because of strong inelasticity of collisions) get attached to
them, $n$ grows until the densely packed state $n=n_c \sim \sigma^{-2}$ is reached. This simple
argument ignores conservation of the total number of particles and cluster edge effects that we
account for in the following.

The vertical force exerted on a particle decreases with an increase of the number of neighboring
particles \cite{Aranson}: the force exerted on a particle located at the cluster edge, $F_e$, is
larger than the force on a particle in the bulk of the cluster $F_b$, but smaller than the force
on a single particle $F_0$. In addition, in a coarse-grained description (which should be valid
for clusters with many particles)  $F_e$ should depend on the local curvature of the cluster edge.
Although  no general relation for $F_e$ is presently available, a clear signature of these cluster
edge effects appears already in the dilute limit where quantitative relations can be obtained.
First, for an inhomogeneous density distribution, the density $n (\mathbf{r})$ in Eq.
(\ref{homogen}) should be replaced by a \textit{locally averaged} density $\overline{n}
(\mathbf{r}) $, the averaging being performed over a region which size is of the order of the
interparticle distance. For a weakly inhomogeneous coverage, one can expand $n(\mathbf{r})$ around
a point $\mathbf{r_0}$ up to $(\mathbf{r}-\mathbf{r_0})^2$. Then, averaging the result over a
circle of radius $\sim n({\mathbf r_0})^{-1/2}$, we obtain $ \overline{n} ({\mathbf r})= n
({\mathbf r})+ {\cal O} (\nabla^2 n/n)$. Substituting it in Eq. (\ref{homogen}), we arrive at
\begin{equation}
F (n, \nabla^2 n)=F_0 (1-k_1\, \sigma^3 n^{3/2}- k_2\,\sigma^3 n^{-1/2} \nabla^2n)\,,
\label{inhomogen}
\end{equation}
where $k_2 = {\cal O}(1)$ is another numerical factor. The cluster edge effects are described by
the $\nabla^2$-term.

Now we start formulating our model. The precipitate number density $n ({\mathbf r},t)$ will serve
as the order parameter, the two phases corresponding to $n=0$ and $n=n_c$ \cite{mono}. In its
turn, the gas density $n_g (t)$ plays the role of the ``mean field". (We assume that the density
relaxation in the gas phase is fast compared to the cluster-gas exchange dynamics. Therefore,
$n_g$ is approximately constant in space and depends only on time.) First, consider a uniform
precipitate in the dilute limit: $n+n_g \ll n_c$. The precipitate will ``evaporate" if $F_0>mg$.
On the contrary, the gas will precipitate if $F_0<mg$. The typical time scales of each of these
two processes are of order of $\tau = \min \left[(g/h)^{1/2},\, \omega^{-1}\right]$. A simple
dynamic equation
\begin{equation}
\frac{dn}{dt}= \frac{\Gamma}{\tau}\left[C\, n_g\theta(\Gamma) +n \,\theta(-\Gamma) \right] ,
\label{eq30}
\end{equation}
subject to conservation law $n+n_g = N/L^2 =const$, accounts for these facts. Here
$\Gamma=1-F_0/mg$ and $\theta(\dots)$ is the $\theta$-function.  The numerical factor $C= {\cal O}
(1)$ accounts for a possible difference between the time scales of ``evaporation" and
precipitation. The factor $\Gamma$ in front of the brackets of Eq. (\ref{eq30}) takes care, in the
simplest possible way, of the continuity of transition between the regimes of $F_0>mg$ and
$F_0<mg$.

For higher precipitate densities one needs to  account for additional effects. First, because of
the screening, the lifting force decreases with $n/n_c$. For simplicity, we assume a linear
dependence $F_0 (1-bn/n_c)$, for all densities $0\le n/n_c \le 1$. Here $b$ is a numerical
constant (using the estimate of $F(n=n_c)$ from Ref. \cite{Aranson}, we have $b \simeq 8/9$).
Second, we account for the inhomogeneity of the precipitate by adopting the $\nabla^2 n$ term from
Eq. (\ref{inhomogen}), with a diffusion coefficient $D \sim \sigma^2/\tau$. Third, we introduce an
additional factor $(n_c-n)/n_c$ in the first term in the r.h.s. of Eq. (\ref{eq30}). It accounts,
in a simple way, for a slowdown of precipitation from the gas phase: at finite $n/n_c$ a part of
the bottom plate is already occupied by grains.

In this way one arrives at a scalar Ginzburg-Landau equation (GLE).
We will be using this equation in the phase coexistence regime $mg<F_0< mg/(1-b)$. (A more detailed
condition will be derived below.) The equation can be written in a scaled form:
\begin{equation}\label{GLE}
  \frac{\partial n}{\partial t} = \phi \,(n, n_g, n_*) + \nabla^2 n\,,
\end{equation}
where
\begin{equation}
\phi=(n-n_*)\times\left\{
\begin{array}{lll}
n,& \mbox{if~}& 0\le n\le n_*\\
Cn_g(1-n),& \mbox{if~}& n_*\le n\le 1  \, .
\end{array}
\right. \label{sh30}
\end{equation}
Here $n_*=(1/b) (1-mg/F_0)$. The coordinates (and the system size $L$, see below) are scaled by
$\delta=(D \tau/\lambda)^{1/2}/n_c$, the time is scaled by $ \tau mg/bF_0$. The precipitate and gas
densities (and $n_*$) are scaled by $n_c$. Finally, $\lambda= bF_0/mgn_c^2$.

The dynamics are constrained by the conservation of the total number of particles.  In the scaled
units
\begin{equation}\label{conservation}
  L^{-2} \int_0^L \int_0^L n(x,y,t) \,dx \,dy + n_g(t) =\varepsilon\,,
\end{equation}
where $\varepsilon$ is the (constant) area fraction of the granulate.

At fixed $n_g$ and $0<n_*<1$, the function $\phi(n,n_g,n_*)$ describes bistability: it has two
stable zeros: at $n=0$ and $n=1$, and an unstable zero at $n=n_*$. Eqs.
(\ref{GLE})-(\ref{conservation}) serve as the continuum version of our model. Similar equations
(with different forms of function $\phi$ and conservation laws) have appeared in other contexts
\cite{Schimansky,Rubinstein,MS,Mikhailov}. The dynamics of a globally conserved bistable system
involve \textit{three} characteristic time scales. Phase separation occurs on the fastest time
scale $t_0$, determined by the properties of the function $\phi$; it is independent of the
characteristic cluster size $l$. On a longer time scale $t_1 \propto l$, the time-dependent gas
density $n_g(t)$ becomes close to special ``equilibrium" value $n_g^{eq}$ such that the area rule
$$
\int_0^1 \phi(n,n_g,n_*) \, dn =0
$$
approximately holds. For the present model, a straightforward calculation yields
\begin{equation}
n_g^{eq}=\frac{n_*^3}{C\,(1-n_*)^3}\,. \label{sh60}
\end{equation}
When $n_g=n_g^{eq}$, a \textit{planar} interface of the precipitate is in a dynamic equilibrium:
it neither advances nor retreats if fluctuations are neglected. For \textit{curved} interfaces
there is an additional, slower \textit{coarsening} stage of the dynamics, governed by the cluster
edge curvature. Its characteristic time $t_2 \propto l^2$ \cite{MS}.

A detailed analysis of this system is possible in the asymptotic \textit{sharp-interface} limit.
The sharp-interface equations become valid towards the end of the $t_1$-stage, when the
precipitate in the clusters is already densely packed ($n=1$), and the gas density $n_g$ is close
to $n_g^{eq}$. In this limit, the conservation law (\ref{conservation}) reads
\begin{equation}
\frac{A (t)}{\hat{L}^2}=\varepsilon -n_g^{eq}\,, \label{conserv}
\end{equation}
where $A (t)$ is the total area of the precipitate. Demanding $A>0$ and using Eq. (\ref{sh60}), we
obtain a condition for the two-phase coexistence:
\begin{equation}
\varepsilon>\frac{n_*^3}{C\, (1-n_*)^3} \,. \label{sh67}
\end{equation}
Using our model expression for $n_*$, we can write
\begin{equation}
1< \frac{F_0}{mg}< \left[1-b\, \left(1+ \frac{1}{\sqrt[3]{C\, \varepsilon}}\right)^{-1}
\right]^{-1}\, . \label{sh360}
\end{equation}
One should compare the more restrictive condition (\ref{sh360}) with the ``na\"{\i}ve''  condition
$1<F_0/mg<(1-b)^{-1}$. For example, for $\varepsilon =0.25, C=1$, and  $b = 8/9$ the
``na\"{\i}ve'' condition gives $1 < F_0/mg < 9$, whereas Eq. (\ref{sh360}) gives $1 < F_0/mg < 1.5
$. The latter condition agrees better with experiment \cite{Aranson}. Even better agreement is
achieved if one chooses $C \simeq 10$.

The normal component of the interface speed is \cite{Mikhailov}:
\begin{equation}
v_n= \nu  C\,\frac{(1-n_*)^3}{n_*^{3/2}}\, (n_g-n_g^{eq}) -{\cal K}\,,
\label{sh110}
\end{equation}
where ${\cal K}$ is the local curvature of the interface
and
$\nu=5(9+2\sqrt{3})/138 \approx 0.45$.

Given the initial location of all interfaces,  Eqs. (\ref{conserv}) and (\ref{sh110}) provide a
proper description of the late-time coarsening dynamics. Furthermore, these equations can be
mapped into equations of {\it interface-controlled} transport which appeared in other contexts.
This enables one to readily present a number of important results for several coarsening
configurations: (1) A planar or circular interfaces are stable with respect to small modulations
\cite{MS,AMS}; (2) The radius of a single cluster which can be in a stable dynamic equilibrium
with the gas phase has a non-zero lower bound. This lower bound scales with the system size (like
$L^{2/3}$), but is independent of $N$ \cite{MS,AMS}; (3) Multiple clusters exhibit Ostwald
ripening: \cite{Schimansky,Rubinstein,MS,AMS}; (4) If there are many clusters, their size
distribution exhibits dynamic scaling. The number of clusters decays with time like $t^{-1}$,
while the average cluster radius grows like $t^{1/2}$ \cite{MS,AMS,W}; (5) Depending on the
initial parameters, a ``hole" (empty region) inside a cluster will either shrink to zero, or
expand and come out of the cluster \cite{MM}. It should be stressed that these results are
insensitive to the exact form of the bistable function $\phi$.

Properties 1, 3 and 4 were already observed in experiment \cite{Aranson}. We checked predictions
1, 2 and 5 by performing additional experiments with cells similar to those described in Ref.
\cite{Aranson}. The experimental results fully agree with these predictions.

\begin{figure}
\centerline{ \epsfig{file=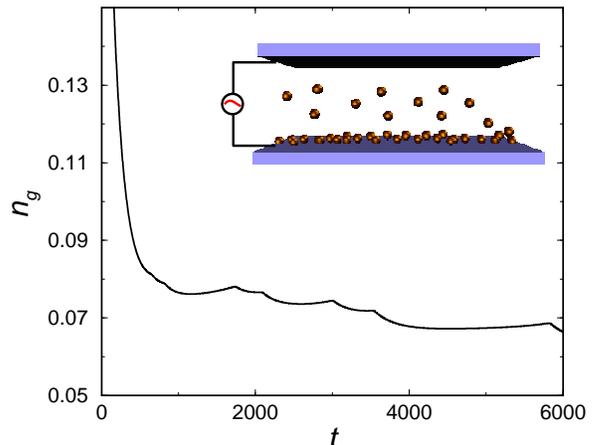, width=3.0in, clip= }}
\caption{The gas phase density $n_g$ (normalized to $n_c$) vs. time. Inset: schematic of the
experimental setting \protect\cite{Aranson}.} \label{fig1}
\end{figure}

Though suitable for theoretical analysis, sharp-interface equations are less convenient for
numerical simulations.
We solved
Eqs. (\ref{GLE})-(\ref{conservation}) numerically in  periodic boundary conditions. Figure 2 shows a
typical simulated evolution of the precipitated phase. The initial conditions describe a situation
when most of the particles are in the low-density precipitate (that is, in the unstable region).
This corresponds to an up-quench of the electric field from $E<E_1$ to the coexistence region
$E_1<E<E_2$. At small times, ``holes" in the precipitate develop and grow. At later times multiple
clusters form and exhibit Ostwald ripening, in agreement with experiment. Figure 1 shows the time
history of the gas density $n_g$ (normalized to $n_c$). At late times, $n_g$ approaches the value
corresponding to the coexistence of a single cluster and gas phase \cite{MS,AMS}. For the set of
parameters chosen for this simulation, the area rule value $n_g^{eq} =0.05$.  The fine structure
of the $n_g (t)$ dependence (small peaks) corresponds to the disappearance of clusters
\cite{AMS,MM}. In experiment, the current through the cell is carried by the grains belonging to
the gas phase, so it should be proportional to $n_g$. Therefore, the ``mean field" $n_g (t)$ is a
directly observable quantity.

The ``up-quench" type of morphology at small times shown in Fig. 2 was not reported in Ref.
\cite{Aranson}, so we performed special experiments with electrostatic cells similar to those used
in Ref. \cite{Aranson}. Starting from the gas phase ($E>E_2$) and making a down-quench to $E<E_1$,
we first prepared a nearly uniform layer of precipitate. Then, after an up-quench into the
coexistence region $E_1<E<E_2$, we observed the same type of morphology as shown in Fig. 2.

\begin{figure}
\centerline{ \epsfig{file=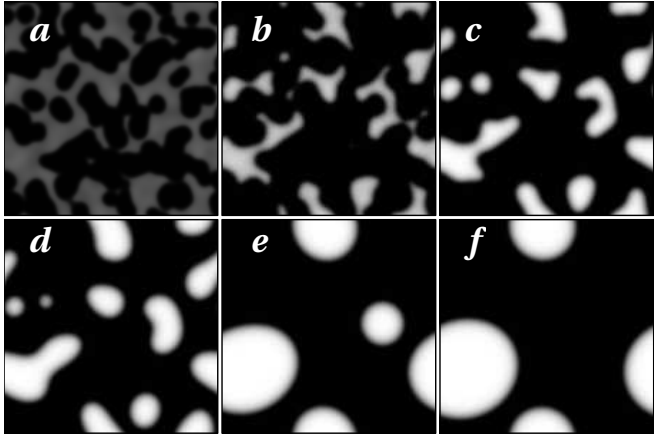, width=3.4in, clip= }} \vspace{0.1in} \caption{Simulated
dynamics of the precipitate for $t=100$ (a), $160$ (b), $300$ (c) $600$ (d) $4000$ (e) and $6000$
(f) scaled time units. White color corresponds to densely packed clusters ($n \simeq n_c$), black
to empty regions.
The system
dimensions (in the units of $\delta$) are $500 \times 500$. The electric field $E$ corresponds to
$n_*= 0.2$.}
 \label{fig2}
\end{figure}
It should be noted that when $F$ is only slightly larger than $mg$, the interface width (which, at
$n_* \ll 1$, is proportional to $n_*^{-3/2}$) becomes comparable to the cluster size and/or
intercluster distance. In this regime the sharp-interface theory is invalid. Still, the continuum
theory should work.

The present model assumes a low-frequency electric field. In this case the gas particles typically
perform an appreciable bounce motion, and a separate, mean-field treatment of the gas phase is
legitimate. For very high frequencies, the particle motion becomes effectively restricted to the
bottom plate. The corresponding coarsening dynamics can be quite different. One can expect that, at
the high frequencies, the $1/2$ growth exponent (observed in experiment \cite{Aranson} and
predicted by our model) will cross over to the $1/3$ growth exponent, typical for
\textit{locally}-conserved systems \cite{LS}.

It is interesting to compare the phase separation properties of electrostatically driven
granulates with those vibrated in the vertical direction mechanically
\cite{Urbach}. Though these two types of systems strongly differ in details of particle
interactions and motion, they have very similar phase diagrams and are strikingly similar in their
phase transition morphologies. This gives an indication that a bistable Ginzburg-Landau equation
subject to conservation of the number of particles can be relevant for the mechanically vibrated
systems as well.

In summary, we have formulated a phenomenological theory of the dynamics of off-equilibrium phase
separation in electrostatically driven conducting monodisperse particles. The theory includes both
continuum, and sharp-interface formulations. It captures the essential physics of this system.
Several new predictions of our model have been verified in experiment.
A better quantitative understanding of this system requires additional quantitative experiments in
different regimes of phase ordering. A comparison of the model with the new experiments will enable
one to determine the unknown numerical coefficients of the model. These coefficients control the
precise phase diagram, the interface velocity, the amplitudes of the dynamic scaling laws in the
Ostwald ripening regime, etc. However, already at this stage we see a strong evidence in favor of
quantitative relevance of the globally constrained GLE to this system.

This research was supported by the US DOE, Office of Basic Energy Sciences, contract \#
W-31-109-ENG-38 and by the Israel Science Foundation, administered by the Israel Academy of
Sciences and Humanities.


\begin{references}
\bibitem{mono} For simplicity, we assume a monolayer coverage, so that $n=n_c$ is the maximum
possible number density of the precipitate.
\end{references}
\end{document}